\documentclass[aps,twocolumn,amsmath,amssymb,superscriptaddress]{revtex4-2}
\pdfoutput=1
\usepackage[colorlinks=true,linkcolor=blue,citecolor=blue,urlcolor=blue]{hyperref}

\usepackage{graphicx}
\usepackage{graphics}
\usepackage[export]{adjustbox}
\usepackage{bm}
\usepackage{amsfonts}
\usepackage{amssymb}
\usepackage{amsmath}
\usepackage{wasysym}
\usepackage{bbold}
\usepackage{stmaryrd}
\usepackage[usenames]{color}
\usepackage{colordvi}
\usepackage{units}
\usepackage{bbm}
\usepackage{booktabs}
\usepackage{array}
\usepackage{placeins}
\usepackage{epsfig}
\usepackage{changes}
\usepackage{braket}
\usepackage{tabularx}
\usepackage[normalem]{ulem}
\newcolumntype{L}[1]{>{\raggedright\arraybackslash}p{#1}} 
\newcolumntype{C}[1]{>{\centering\arraybackslash}p{#1}} 
\newcolumntype{R}[1]{>{\raggedleft\arraybackslash}p{#1}} 

\newcommand{\be}{\begin{equation}}
\newcommand{\ee}{\end{equation}}
\newcommand{\beqn}{\begin{eqnarray}}
\newcommand{\eeqn}{\end{eqnarray}}

\usepackage{orcidlink} 
 
\bibstyle{apsrev.bib}

\begin{document}

\title{Quantum Potts chain in alternating field}
\author{P\'eter Lajk\'o}
\email{peter.lajko@ku.edu.kw}
\affiliation{Department of Physics, Kuwait University, P.O. Box 5969, Safat 13060, Kuwait}
\author{Wedade Alaaeldin Ahmed Shafik Yehia}
\affiliation{Department of Physics, Kuwait University, P.O. Box 5969, Safat 13060, Kuwait}
\author{Ferenc Igl{\'o}i}
\email{igloi.ferenc@wigner.hu}
\affiliation{Wigner Research Centre for Physics, Institute for Solid State Physics and Optics, H-1525 Budapest, P.O. Box 49, Hungary}
\affiliation{Institute of Theoretical Physics, Szeged University, H-6720 Szeged, Hungary}
\date{\today}

\begin{abstract}
The $q$-state Potts chain with ferromagnetic couplings, $J=1$, in the presence of a transverse field, $\Gamma$, has a quantum phase transition at $\Gamma/q=1$, which is continuous for $q \le 4$ and of first order for $q>4$. Here we introduce a $q$-periodic alternating longitudinal field of strength, $h$, and study the phase diagram and the critical properties of the model. For $h<q/(q-1)$ there is a ferromagnetic ordered phase, for $\Gamma<\Gamma_c(h)$ and at $h=q/(q-1)$ there is a classical endpoint at $\Gamma=0$, with finite entropy at $T=0$. We considered the $q=3$ model and using DMRG techniques we calculated the low-laying spectrum of the Hamiltonian, the transverse magnetisation and the spin-spin correlation function, all of which signalled a diverging correlation length at the transition point with the exponent of the three-state Potts model. In the vicinity of the classical endpoint the model is mapped to a quantum hard rod model, which belongs also to the universality class of the three-state Potts model. Also the spectrum of the critical Hamiltonian is found in agreement with conformal invariance. At the same time the correlation function shows a jump at the transition point, thus the transition is of mixed order for $h<q/(q-1)$.

\end{abstract}


\maketitle

\section{Introduction}
\label{sec:intro}
Quantum phase transitions takes place at zero temperature due to quantum fluctuations by varying a control parameter, such as the strength of a transverse field, $\Gamma$\cite{sachdev_2011}. In real materials the influence of a quantum phase transition can be noticed by the specific behaviour in the low-temperature behaviour. Due to the nature of quantum fluctuations the critical singularities at a $d$-dimensional quantum system are related to that in a classical ($d+1$)-dimensional system. Therefore quantum phase transitions takes place in $d=1$, too, which systems can be well experimentally realized by ultracold atomic gases in optical lattices. One such basic example is the antiferromagnetic Ising chain in the presence of transverse and longitudinal fields defined by the Hamiltonian:
\begin{align}
 \hat{H}_I&=J_I\sum_{i=1}^L \sigma_{i}^{z} \sigma_{i+1}^{z}
-\Gamma_I\sum_{i=1}^L  \sigma^x_{i} -h_I\sum_{i=1}^L \sigma^z_{i}\;.
\label{Hamilton_I}
\end{align}
in terms of the $\sigma_{i}^{x,z}$ Pauli matrices at site $i$. In the experiments the longitudinal fields are generally not too strong\cite{Simon2011}.

On the theoretical side, the model in Eq.(\ref{Hamilton_I}) has been studied by different methods: finite-size exact diagonalization\cite{PhysRevE.63.016112}, density matrix renormalization (DMRG)\cite{PhysRevB.68.214406,PhysRevB.101.024203,PhysRevB.103.174404}, quantum Monte Carlo simulation\cite{PhysRevB.96.064427}, real-space renormalization group\cite{PhysRevB.103.174404}, the fidelity susceptibility\cite{PhysRevE.99.012122} and the pattern description\cite{yang2023} methods. The quantum phase transition in the model is found of mixed-order\cite{PhysRevB.103.174404}, which means a diverging correlation length and a finite jump of the correlation function at the transition point.

The model in Eq.(\ref{Hamilton_I}) can be generalised in different ways. First we consider $m$-spin interactions, in which case the Hamiltonian reads as\cite{PhysRevB.37.7884}:
\begin{align}
\begin{split}
\hat{H}_M&=(-1)^m J_M\sum_{i=1}^L \prod_{j=1}^{m}\sigma_{i+j-1}^{z}\\
&-\Gamma_M\sum_{i=1}^L  \sigma^x_{i} -h_M\sum_{i=1}^L \sigma^z_{i}\;.
\label{Hamilton_M}
\end{split}
\end{align}
In the absence of longitudinal field, $h=0$, (which is the Turban model\cite{LTurban_1982}), the model has a quantum phase transition at $\Gamma_M/J_M=1$, at which point the degeneracy of the ground state is lifted by a factor of $2^{m-1}$. The transition is continuous for $m=3$, which belongs to the 4-state Potts universality class\cite{LTurban_1982,FIgloi_1983,CVanderzande_1987,FIgloi_1987} and it is of first order for $m \ge 4$\cite{FIgloi_1986}. For negative longitudinal fields, $h_M<0$, the degeneracy at the quantum phase-transition line is lifted by a factor $m$ and for $m=3$ it is found numerically to belong to the 3-state Potts universality class\cite{PhysRevB.37.7884}. The phase-transition line terminates at a classical endpoint (CEP) at $h_M=-mJ_M$ and $\Gamma_M=0$, where the system has finite entropy per site. In the vicinity of the CEP for $h_M+mJ_M \to 0^+$ and $\Gamma \to 0$ the system is effectively described by a so called quantum hard rod model\cite{PhysRevB.40.2362}.

Here we introduce another generalization of the Hamiltonian in Eq.(\ref{Hamilton_I}), when the Ising spins are replaced by $q$-state Potts variables and the longitudinal field has a $q$-periodic alternation:
\begin{align}
\begin{split}
\hat{H}&=-J\sum_{i=1}^L \delta_{s_{i},s_{i+1}}\\
&-\Gamma\sum_{i=1}^L \sum_{k=1}^{q-1}M_i^k  -h\sum_{i=1}^L \delta_{s_{i},(i-1,{\rm mod}~q)}\;.
\label{Hamilton_P}
\end{split}
\end{align}
Here $s_i=0,1,\dots,q-1$ denotes a Potts spin variable at site $i$, $M_i$ is a spin-flip operator: $M_i^k |s_i\rangle=|s_i+k,{\rm mod}~q\rangle$, see in\cite{PhysRevB.24.218}. $J=1$ is the ferromagnetic coupling, $\Gamma$ and $h$ are the strengths of the transverse and the alternating longitudinal field, respectively. Indeed for $q=2$ the Hamiltonian in Eq.(\ref{Hamilton_I}) is equivalent to that in Eq.(\ref{Hamilton_P}) after a gauge transformation: $\sigma_i^z=(-1)^{i}\sigma_i^z$.

In this paper we aim to study the properties of the quantum phase transitions in this generalized quantum Potts model, which is expected to show some similarities with the phase transitions in the multispin coupling model for $h_M<0$. Having $q=m$ the degeneracy of the ground state in the two models is lifted by the same factor at the phase-transition line and the classical models (with zero transverse fields) have similar phase structure. In particular we will concentrate on the $q=3$ model, which will be investigated numerically by the DMRG method. We study the critical behaviour of the spin-spin correlation function, the scaling behaviour of the transverse magnetisation, as well as the finite-size scaling behaviour of the spectrum at the critical point and its relation to conformal invariance.

The rest of the paper is organised in the following way. The phase diagram of the model is calculated in Sec.\ref{sec:phase_diag} using different numerical methods. In Sec. \ref{sec:critical}, the critical properties are calculated, when the transition line is approached from the ordered phase ($\Gamma \le \Gamma_c$) and from the alternating phase ($\Gamma > \Gamma_c$). The results are discussed in Sec.\ref{sec:disc}, while a mapping of the model in the vicinity of the critical end-point is presented in the Appendix.

\section{Phase diagram}
\label{sec:phase_diag}

We have studied the properties of the $q=3$ model numerically by the DMRG method. We used finite samples of length $L=6\ell$, $\ell=1,2,\dots$ so that a complete period of the longitudinal field fits to the half of the chain. We went up to $L=96$ and the ground state and the first few excited states are calculated. Here the original version of the infinite system DMRG scheme was utilized\cite{PhysRevLett.69.2863,PhysRevB.48.10345} with periodic boundary conditions. For the correlations it was systematically and carefully checked that their values are independent of the basis size $\mu$ within the error bars. The accuracy of the ground-state energy calculations was in the range $10^{-6}-10^{-8}$ and this was in full agreement with the truncation error, the largest basis size being $\mu=244$ for the different systems.
\begin{figure}[h!]
\begin{center}
\includegraphics[width=\linewidth,angle=0]{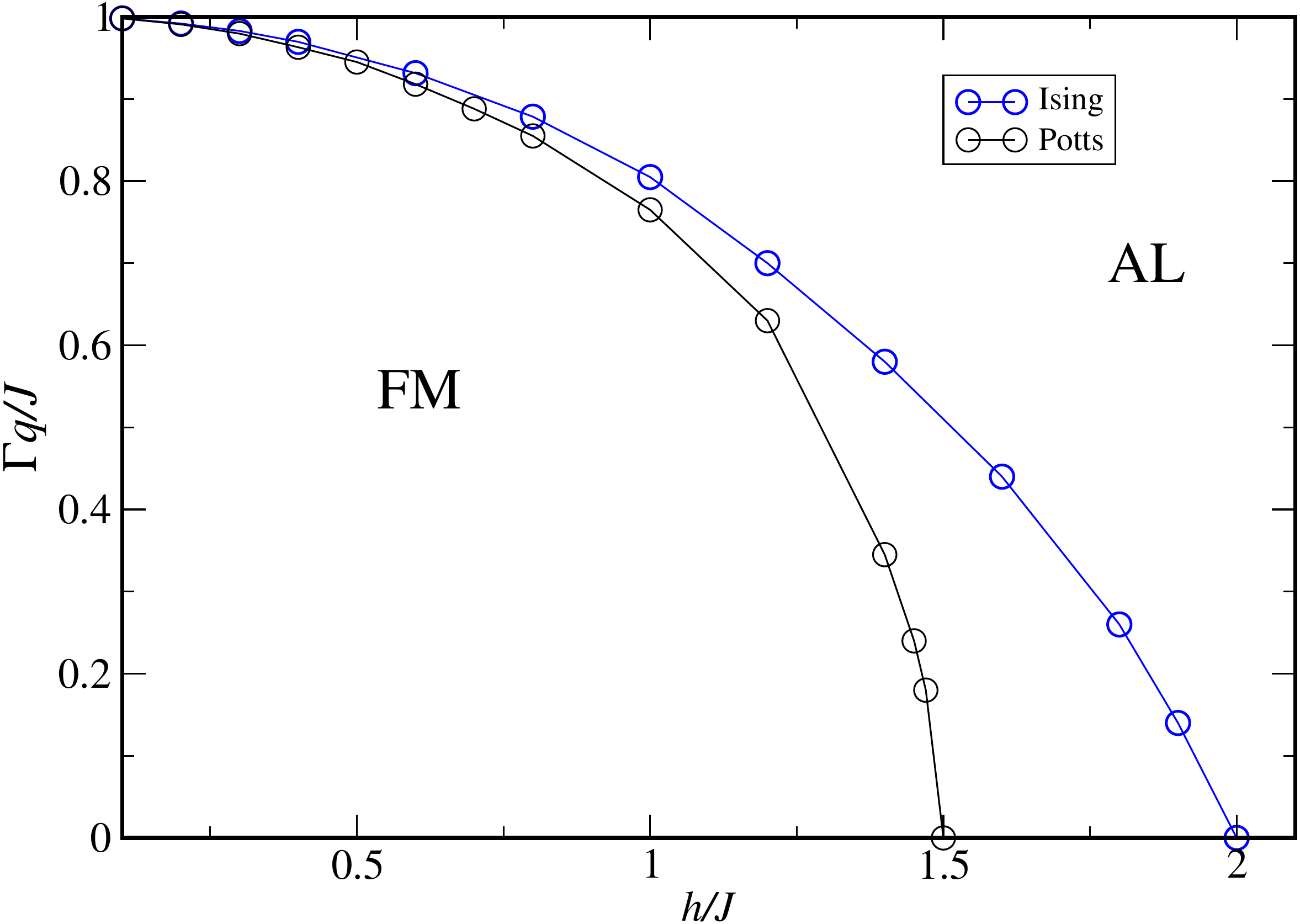}
\end{center}
\caption{\label{fig:fig1}(Color online) Zero-temperature phase diagram of the Potts chain for $q=3$ (black symbols) with ferromagnetic coupling, $J$,
in transverse ($\Gamma$) and alternating ($h$) magnetic fields. The phase diagram of the Ising chain ($q=2$ blue symbols) is also shown for comparison.
In the absence of alternating field, $h=0$, the transition between a quantum ferromagnetic (FM) phase and a quantum paramagnetic (PM) phase is controlled by the fixed point of the quantum Potts model (QPM) at $(\Gamma q/J=1,\,h/J=0)$. For finite value of $h>0$ the FM phase survives and at the other side of the phase boundary there is an alternating phase (AL). In the classical limit, $\Gamma=0$,  there is a classical end-point (CEP) at $(h/J=q/(q-1))$, having a first-order transition.
}
\vskip-3mm
\end{figure}

The calculated phase diagram is shown in Fig.\ref{fig:fig1} for $q=3$, which is compared with that of the Ising case, $q=2$. The phase diagram of the two systems has the same topology and this is expected to hold for larger values of $q$, too. There is a ferromagnetically ordered phase for $\Gamma<\Gamma_c(h)$ in the region $0 \le h < h_c$. At the two extreme points of the phase boundary there are exact results, which hold for general values of $q$. In the absence of alternating longitudinal fields, $h=0$, the quantum phase transition is located at $\Gamma_c(0)=qJ$\cite{PhysRevB.24.218}, which is continuous for $q \le 4$ and of first order for $q>4$\cite{RJBaxter_1973}. For $q=3$ the critical exponents are known through conformal invariance\cite{DOTSENKO1984312} and Coulomb-gas mapping\cite{MPMdenNijs_1979,PhysRevB.22.2579,PhysRevB.27.1674} as $\nu=5/6$ and $\beta=1/9$ and the central charge is $c=4/5$. 

The other end of the phase-boundary is located at $\Gamma_c=0$, where there is a classical end point (CEP) at $h_c=Jq/(q-1)$. Here the ferromagnetic and the alternating phases meet and the transition is of first order. At the CEP the ground state is infinitely degenerate, the entropy per spin is finite. In the vicinity of the CEP the phase boundary in terms of $h_c-h=\Delta h>0$ and $\Gamma>0$ is linear for $q=2$, $\Gamma_c \sim \Delta h$, but for $q=3$ it looks quadratic: $\Gamma_c^2 \sim \Delta h$. This result will be explained in the Appendix, by using the fact that the degeneracy of the ground state is solved up and in leading order the problem is mapped to a set of secular equations, which involve the degenerate states. For general value of $q$ the phase boundary at the CEP is expected to behave as $\Gamma_c^{q-1} \sim \Delta h$.
\begin{figure}[h!]
	\begin{center}
		\includegraphics[width=8.6cm,angle=0]{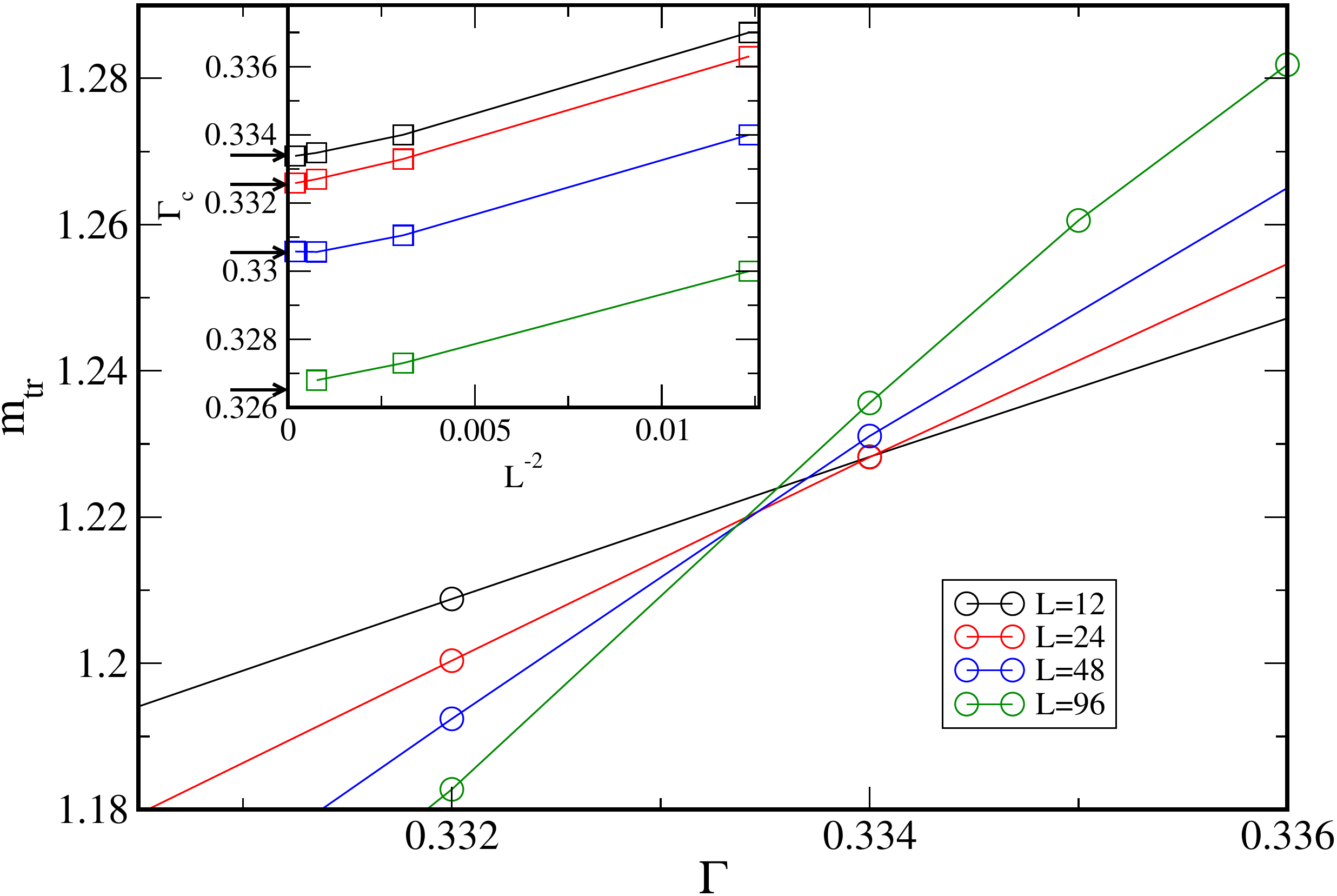}
	\end{center}
	\caption{\label{fig:fig2}(Color online) The transverse magnetisation, $m_{tr}$, as a function of $\Gamma$ for different finite systems with $L=12, 24, 48, 96$ at $h=0$. The crossing points of the curves define finite-size transition points. Inset: extrapolation of the finite-size transition points as $L^{-2}$ for different values of $h=0.0, 0.1, 0.2$ and $0.3$, up to down. The arrows indicate the extrapolated values: $\Gamma_c=0.3327, 0.3306, 0.3265$ at $h=0.1, 0.2, 0.3$, respectively.}
\end{figure}

For the $q=3$ model the phase-transition line for general values of $0<h<h_c=3J/2$ has been estimated by different methods. One procedure is based on the transverse magnetisation, which is defined as:
\be
m_{tr}=\frac{1}{L}\sum_{i=1}^L \left\langle \sum_{k=1}^{q-1}M_i^k \right\rangle\;.
\ee
Having a fixed value of $h$, $m_{tr}$ is plotted in Fig.\ref{fig:fig2} as a function of $\Gamma$ for different finite systems. It is seen that the $m_{tr}$ is a monotonously increasing function of $\Gamma$ and the slope of the curves increases with the size, $L$. Consequently the curves cross for different $L$ values in the vicinity of the transition point. Such a crossing point, with $m_{tr}(\Gamma,L_1)=m_{tr}(\Gamma,L_2)$ is identified as a finite-size transition point: $\Gamma_c(L_1,L_2)=\Gamma$. With increasing values of $L_1$ and $L_2$, $\Gamma_c(L_1,L_2)$ approach the true transition point, as illustrated in the inset of Fig.\ref{fig:fig2}.

\begin{figure}[t!]
	\begin{center}
	\includegraphics[width=7.6cm,angle=0]{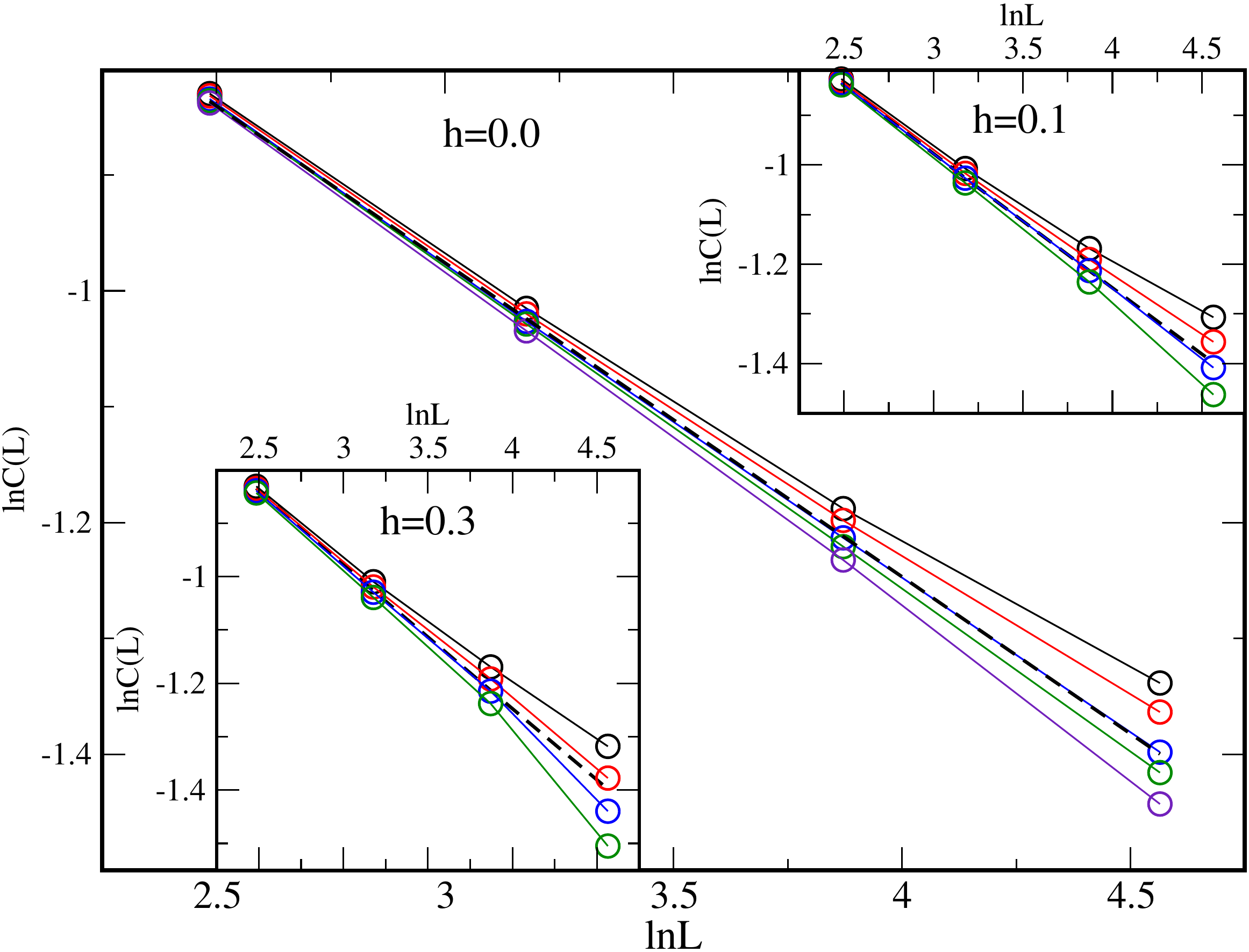}
	\end{center}
	\caption{\label{fig:fig3}(Color online) Log-log plots of the correlations for $h=0.0$, from up to down $\Gamma=0.3331, 0.3332, 1/3, 0.3334, 0.3335$. The dashed line is the fit of the points at $\Gamma=1/3$. Upper inset: the same for $h=0.1$, from up to down $\Gamma=0.3322, 0.3324, 0.3326, 0.3328$. The dashed line is taken from the main figure. Lower inset: the same for $h=0.3$, from up to down $\Gamma=0.3262, 0.3264, 0.3265, 0.3266, 0.3268$. The dashed line is taken from the main figure.}
\end{figure}

Somewhat more accurate estimates have been obtained through the spin-spin correlation function, which is defined as:
\be
C(i,i+r)=\frac{q \langle \delta_{s_{i},s_{i+r}}\rangle-1}{q-1}\;.
\label{C_r}
\ee
In the thermodynamic limit, $L \to \infty$ and for $r \to \infty$ the correlation function has the following limiting values. In the ferromagnetic phase $0<C(i,i+r)\le 1$, in the paramagnetic phase, $C(i,i+r)=0$  and in the alternating phase, for $r \ne qn$, $n=1,2,\dots$ it is $-0.5 \le C(i,i+r)< 0$. For a finite chain we consider the largest, non-periodic separation and use $C(L)\equiv C(1,L/2)$ in the investigations. To locate the phase-boundary we make use of the fact that at a second-order transition point the correlation function decays as a power-law with the distance:
\be
C(L) \sim L^{-\eta}\;,
\ee
where the decay exponent satisfies the relation: $\eta=2 \beta/\nu$. To check this relation we plot in Fig.\ref{fig:fig3} for a few values of $h=0.0$, $0.1$ and $0.3$, the correlation function, $C(L)$ vs. $L$ in double-logarithmic scale for different values of $\Gamma$, close to the expected phase-transition point. It is seen in this figure, that for $h=0$, this type of analysis reproduces very accurately (up to four digits) the exactly known phase-transition point and the slope of the straight line coincides with the analytical result: $\eta=4/15$. Similar analysis for other values of $h>0$ results in comparable precision both with the location of the transition point and the slope of straight line.

The phase-boundary can also be determined through the use of the scaled gap:
\be
F(\Gamma,L)=(E_1(\Gamma,L)-E_0(\Gamma,L))L\;.
\ee
where $E_0$ and $E_1$ denotes the energy of the ground state and the first excited state, respectively. For large systems 
$F(\Gamma,L) \to \infty$ in the alternating phase and $F(\Gamma,L) \to 0$ in the ferromagnetic phase. The phase-transition point separates these two regions, where $F(\Gamma,L)$ goes to a constant, the value of which is determined through conformal invariance.
By this method we didn't make a detailed analysis, just checked, that at the critical point, which was earlier numerically calculated the scaled gap indeed satisfies the expected scaling properties (see the calculations around Eq.(\ref{gap})).

\section{Critical properties}
\label{sec:critical}

We have studied the singular behaviour of different physical quantities in the vicinity of the phase-transition point for relatively small values of the alternating longitudinal field, $h \le 0.3$, and compared it with the results obtained in the absence of longitudinal fields. We note, that the circumstances in the two cases are not completely equivalent: for $h=0$ in the strong transverse-field limit, $\Gamma>\Gamma_c$, the system is paramagnetic and $\lim_{L \to \infty} C(L)=0$, whereas for $h>0$ there is an alternating phase, with $\lim_{L \to \infty} C(L)<0$. Therefore we consider first the weak transverse-field region, with $\Gamma \le \Gamma_c$, and we treat the strong transverse-field region afterwards.

\subsection{Weak transverse-field region: $\Gamma \le \Gamma_c$}

We start to discuss the behaviour of the spin-spin correlation function at the critical point, $\Gamma=\Gamma_c$, which has been illustrated in Fig.\ref{fig:fig3}. This critical correlation function is found to have a power-law decay and the decay exponents are found to be independent of the strength of the alternating longitudinal field, their values agreed with the corresponding value for $h=0$, $\eta=4/15$, within the error of the calculation. 

Next we turn to discuss the behaviour of the transverse magnetisation, $m_{tr}$, which was illustrated in Fig.\ref{fig:fig2}. According to scaling theory in the vicinity of the critical point $m_{tr}$ is divided into a non-fluctuating part, $m_{tr}^*=m_{tr}(\Gamma_c)$, and a fluctuating part, $\tilde{m}_{tr}$. The latter scales in the same way as the energy-density: $\tilde{m}_{tr}(\Gamma) \approx ({\rm d}\tilde{m}_{tr}/{\rm d} \Gamma) \Delta \Gamma \sim (\Delta \Gamma)^{1-\alpha}$, where $\Delta \Gamma=\Gamma_c-\Gamma$ and $\alpha$ is the critical exponent of the specific heat. In a finite system the correlation length, $\xi$, is limited by the linear extension of the system, thus $L \sim \xi \sim (\Delta \Gamma)^{-\nu}$, where $\nu$ is the correlation-length critical exponent. Consequently the slope of the transverse magnetisation is given by:
\be
\frac{{\rm d}\tilde{m}_{tr}}{{\rm d} \Gamma} \sim (\Delta \Gamma)^{-\alpha} \sim L^{\omega}\;,
\label{alpha/nu}
\ee
with $\omega=\alpha/\nu$. Using the hyperscaling relation\cite{cardy_1996} for a one-dimensional quantum system: $(1+1)\nu=2-\alpha$, we obtain for the exponent in Eq.(\ref{alpha/nu}) $\omega=2(1-\nu)/\nu$, or $1/\nu=1+\omega/2$.

\begin{figure}[h!]
	\begin{center}			
		\includegraphics[width=8.6cm,angle=0]{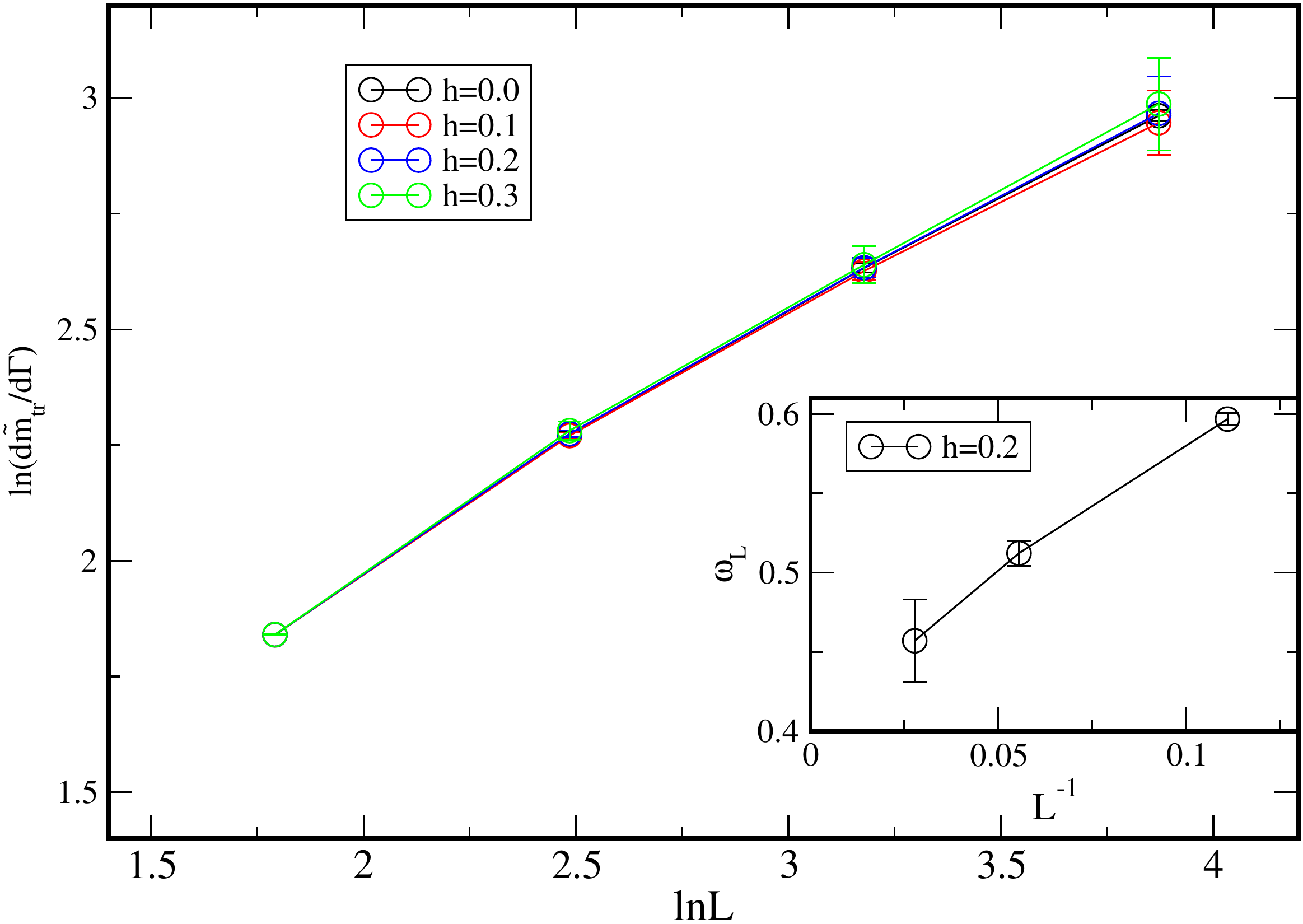}
	\end{center}
	\caption{\label{fig:fig4}(Color online) Derivative of the transverse magnetisation at the critical point as a function of the length of the chain in a log-log scale for different values of $h=0.0,0.1,0.2$ and $0.3$. In the inset size-dependent effective exponents, $\omega_{L}$ are presented, which are calculated from the local slopes of the curves in the main figure for $h=0.2$. Extrapolation is performed by assuming finite-size corrections in the form $1/L$.}
\end{figure}

The measured derivatives of the transverse-magnetisation at the critical point are shown in Fig.\ref{fig:fig4} as a function $L$ in log-log scale. The points for different values of $h$ practically overlap and from the local slope of the curve estimates for the effective, size-dependent exponent $\omega_L$ can be obtained, which are shown in the inset of Fig.\ref{fig:fig4} vs. $1/L$ for $h=0.2$. The extrapolated value for $L \to \infty$ is $\omega=0.40(3)$, from which the correlation-length exponent is given by $1/\nu=1.20(2)$, in agreement with the value of the three-state Potts model.

We have also estimated finite-size scaling behaviour of the ground-state energy, $E_0(L)$, and that of the first gap, $\Delta E(L)=E_1(L)-E_0(L)$ at the critical point. According to conformal invariance for periodic chains $E_0(L)$ asymptotically behaves for large-$L$ as\cite{PhysRevLett.56.742,PhysRevLett.56.746}:
\be
\frac{E_0(L)}{L}=e_0-\frac{\pi v}{6 L^2}c\;,
\label{casimir}
\ee
where $c$ is the central charge of the Virasoro-algebra and $v$ denotes the so-called sound velocity. For two finite systems of lengths, $L_1$ and $L_2$ we obtain the following estimate for the prefactor of the correction term:
\be
{[E_0(L_1)/L_1-E_0(L_2)/L_2]\frac{L_1^2L_2^2}{L_2^2-L_1^2}}=\frac{\pi}{6 }vc(L_1,L_2).
\label{vc}
\ee
The finite-size scaling form of the energy gap at the critical point is given by\cite{JLCardy_1984,GvonGehlen_1986,cardy_1996}:
\be
\Delta E(L)=\frac{2\pi v}{L} x\;,
\label{gap}
\ee
with $x=\eta/2$ being the scaling dimension of the magnetisation. Here we have calculated the average of the prefactors for two chains with lengths $L_1$ and $L_2$ as:
\be
{(L_1\Delta E(L_1)+L_2\Delta E(L_2))/2}=2\pi vx(L_1,L_2).
\label{vx}
\ee
The ratio of the two expressions in Eqs.(\ref{vc}) and (\ref{vx}) is
\be
r_{L_1,L_2}=\frac{1}{12}\frac{c(L_1,L_2)}{x(L_1,L_2)},
\label{ratio}
\ee
 which should tend to $1/2$ for the three state Potts model, since the central charge is $c=4/5$. Numerical results for this ratio are shown in Table \ref{table1} for the model with an alternating field of strength $h=0.3$ and for comparison we present the results for the case, $h=0$, too. Note, that here results for relatively small chains are presented, however with large numerical precision, which is needed to obtain correct estimates even for the correction terms. The numerical estimates for the ratio are fairly close to the expected asymptotic value and similar trend is seen for the standard model with $h=0$.

\begin{table}[h!]
\begin{center}
\begin{tabular}{||c | c | c ||} 
 \hline
 $L_1,L_2$ & $r_{L_1,L_2}^{(h=0.3)}$& $r_{L_1,L_2}^{(h=0.)}$ \\ [0.5ex] 
 \hline\hline
 6,12& 0.4997 & 0.4993  \\ 
 \hline
 12,18&0.4981 &0.4982  \\
 \hline
 18,24 & 0.4978 &0.4983  \\
 \hline
 24,30 & 0.4983 & 0.4984  \\
 \hline
\end{tabular}
\end{center}
\caption{\label{table1}Estimates for the ratio defined in Eq.(\ref{ratio}) for the model with an alternating field of strength $h=0.3$ and for the standard model with $h=0$.}
\end{table}

Next we turn to analyse the behaviour of the system in the vicinity of the CEP point in the limit $\Delta h=h_c-h \ll 1$ and $\Gamma \ll 1$. Here the effective Hamiltonian is spanned by the states that are degenerate ground states at the CEP and whose degeneracy is resolved by switching on the fields. We observed and explained in the Appendix, that this effective Hamiltonian is related to a similar effective Hamiltonian of the multispin-coupling model in Eq.(\ref{Hamilton_M}). As a matter of fact there is a direct mapping between the degenerate ground states in the CEP points of the two models. In the multispin-coupling model we obtain a set of secular equations in first-order degenerate perturbation calculation, which is equivalent to a quantum-hard-rod problem\cite{PhysRevB.40.2362}. For the $q=3$ model the perturbation needs to be treated in second order, which explains the observed quadratic behaviour of the phase-boundary at the CEP in Fig.\ref{fig:fig1}. As a further consequence the critical behaviour of the $q$-state quantum Potts chain at the CEP is identical, to that of the quantum-hard-rod model for $q=m$. The $m=3$ model has been studied some time ago. Although earlier studies on relatively smaller systems are in favour of a first-order transition\cite{PhysRevB.40.2362,PhysRevB.41.4475}, Monte Carlo simulations on special Ising simulator found second-order transition, with exponents in the three-state Potts universality class\cite{FIgloi_1992}. This result should hold also for the $3$-state Potts model with $h>0$ alternating longitudinal field.

\subsection{Strong transverse-field region: $\Gamma > \Gamma_c$}

\begin{figure}[h!]
	\begin{center}
		\includegraphics[width=8.6cm]{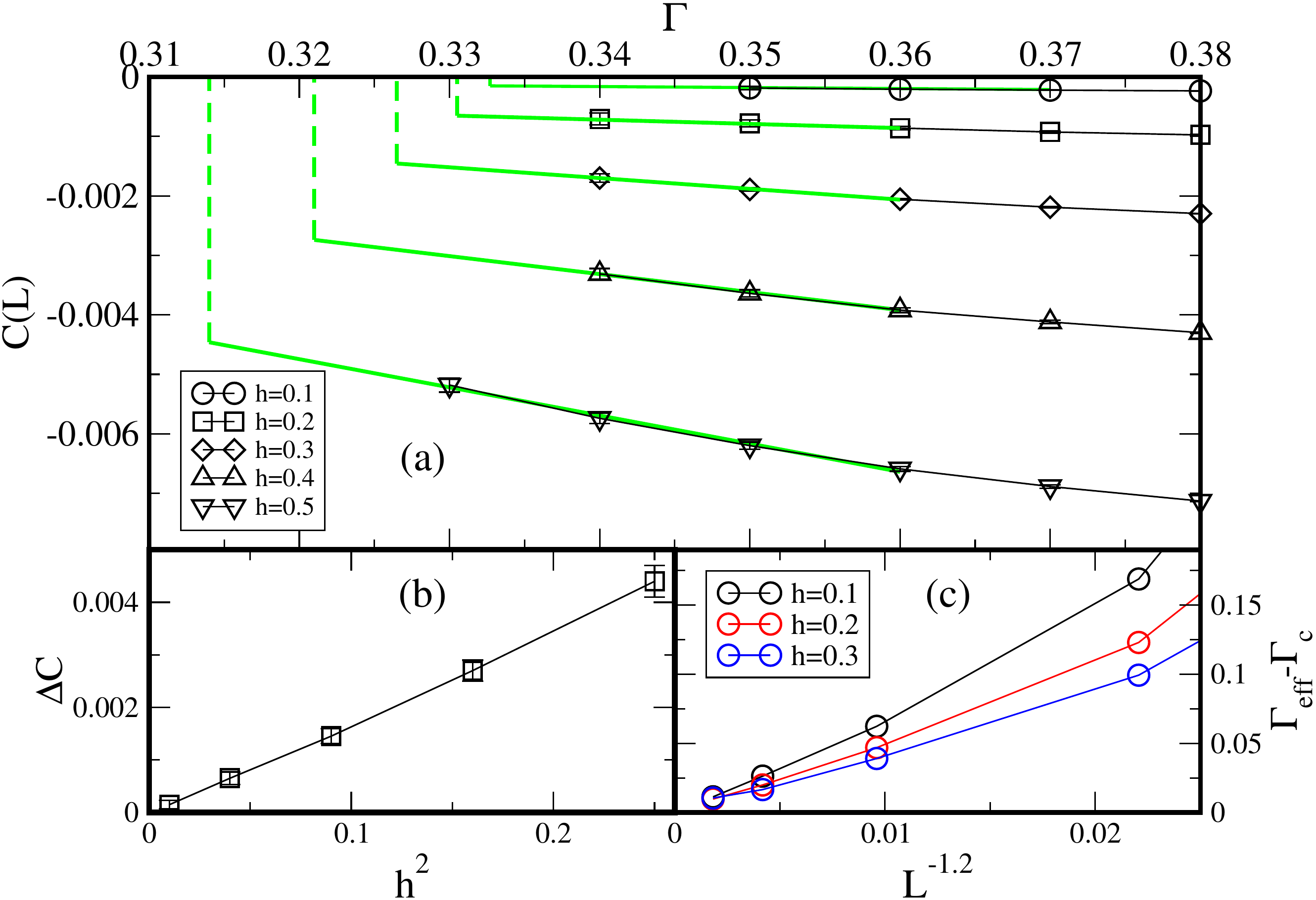}
	\end{center}
	\caption{\label{fig:fig5}(Color online)  The correlation function just above the critical point as a function of $\Gamma>\Gamma_c$ for different values of $h=0.1$, $0.2$, $0.3$, $0.4$, and $0.5$, from top to bottom. Green solid lines are fits to the correlation data and also serve as guide to the eye, they end at the location of the phase transition, which was calculated in Sec.\ref{sec:phase_diag}. Inset (b):  The estimated jumps in the correlation function at the critical point as a function of $h^2$. Inset (c): The difference between the value of $\Gamma_{0}(L)$ where the sign change of the correlation function takes place and the value of $\Gamma_c$ as a function of $L^{-1/\nu}$, with $\nu=5/6$.}
\end{figure}

In the strong transverse-field region: $\Gamma > \Gamma_c$, the spin-spin correlation function is $C(L)<0$, and monotonously increases with decreasing $\Gamma$. This is illustrated in Fig.\ref{fig:fig5}. Here the calculated points have practically no size-dependence. If we approach the phase-transition point closer, we can not exclude finite-size effects, which are enormously large in the vicinity of the transition points. These transition points are calculated in Sec.\ref{sec:phase_diag} and these are indicated by vertical dashed lines. If we assume, that the correlation function close to the transition point can be obtained by analytically continuing the curves, which go through the calculated points we obtain the continuation of the curves, which are indicated in Fig.\ref{fig:fig5}(a). This construction results in a discontinuous jump of the correlation function. The estimated values of the jumps have an approximately quadratic dependence: $\Delta C \sim h^2$, as shown in the Fig.\ref{fig:fig5}(b). We mention, that similar behaviour was observed in the $q=2$ Ising case, too\cite{PhysRevB.103.174404}. 

At the same time, however, the correlation length stays divergent also for $\Gamma - \Gamma_c \to 0^+$. This can be seen, if we compare the position, $\Gamma_0(L)$, where $C(L)=0$ with the phase-transition point, $\Gamma_c$. In Fig.\ref{fig:fig5}(c) we see that this difference scales with the size as $\Gamma_0-\Gamma_c \sim L^{-1/\nu}$, with $\nu=5/6$ being the correlation-length critical exponent.

\section{Discussion}
\label{sec:disc}
The $q$-state Potts model is a basic system of statistical mechanics the properties of its phase transition have been thoroughly studied by different methods\cite{RevModPhys.54.235}. Here we considered the one-dimensional model in a transverse field, which is the quantum Potts chain\cite{PhysRevB.24.218} the critical properties being identical to that of the two-dimensional classical model. In our study the Hamiltonian is extended with an alternating longitudinal field, which perturbation will keep the ferromagnetic order in the system, at least for not too large transverse and longitudinal fields. On the other hand the inclusion of the alternating field  will modify the properties of the non-ordered phase, what we call alternating phase. In the alternating phase the spin-spin correlation function has a finite negative value, whereas in the paramagnetic phase, which is the case in the standard model (without an alternating field) the spin-spin correlation function vanishes in the thermodynamic limit. We have studied the properties of the phase transition in the system between the ferromagnetic and the alternating phases. In details we considered the $q=3$ state model and calculated different properties (transverse magnetisation, spin-spin correlation function, finite-size scaling of the gap) by the DMRG method. At the phase-transition line we observed singular behaviour of the different physical quantities which are connected to a diverging correlation length. The calculated values of the critical exponents (correlation-length exponent, magnetization exponent, central charge) are found consistent with the values of the three-state Potts model. This type of universality is further supported by the analysis of the phase-transition in the vicinity of the critical end-point, which is located at $\Gamma=0$ and $h_c=q/(q-1)$. Here the transition line starts as $h_c-h \sim \Gamma^{q-1}$ and the model can be mapped to a so called quantum hard rod model\cite{PhysRevB.40.2362}, with rod-length $m=q$. Earlier studies of the $m=3$ quantum rod model also predicted the transition in the $q=3$-state Potts universality class.

New insight about the transition is found if the transition line is approached from the side of the alternating phase. Considering the spin-spin correlation function it is negative in the alternating phase and our numerical results indicated, that it vanishes discontinuously at the transition point. Therefore we conjecture that the transition is of mixed order. 
Mixed-order transitions have already been observed in different systems: in clssical spin chains with power-law type long-range interactions\cite{PhysRevLett.23.89,PhysRev.187.732,Dyson1971,JLCardy_1981,Aizenman1988,SLURINK1983627,PhysRevLett.112.015701,PhysRevE.94.062126}, in models of depinning transition\cite{ 10.1063/1.1727785,10.1063/1.1727787,PhysRevE.52.1223,Fisher1984}, in percolation models with glass and jamming transition\cite{PhysRevLett.55.304,PhysRevLett.96.035702,PhysRevLett.98.129602,J.M.Schwarz_2006,PhysRevLett.109.205703,Liu_2012,ZIA2012124,TIAN2012286,PhysRevE.86.011128,PhysRevLett.114.098104}, in models at amultiple-junction or $k$-booklet geometry\cite{FIgloi_1991,JLCardy_1991,PhysRevE.95.022109,PhysRevE.95.010102}, in modular networks\cite{PhysRevResearch.3.013106}, and in experiments\cite{doi:10.1073/pnas.1712584114}, for a recent review see\cite{Bar_2014,mukamel2023}.
Mixed-order transition has already been identified for the $q=2$ model, which is just the Ising model and the same phenomenon seems to be valid for the $q=3$ model, too\cite{PhysRevB.103.174404}.

In the Introduction we have also presented the multispin-coupling model in Eq.(\ref{Hamilton_M}), which contains an $m$-spin product form and for negative longitudinal field $h_M<0$ shows similar critical properties as the Potts model in an alternating field for $q=m$. First, the degeneracy of the ordered phase is lifted by the same factor, $q=m$ in the two models and in the vicinity of the critical end-points there is an exact mapping between the models through the hard rod model in Eq.(\ref{rod}). Numerical studies of the $q=3$ model and that of the $m=3$-coupling model show critical exponents in the three-state Potts universality class. Since the $q=3$ model shows a mixed-order transition the same type of scenario could be valid for the $m=3$-coupled Ising model, too. To check this conjecture further investigations are necessary.

\section*{Appendix: Effective Hamiltonian at the critical end-point - relation with the quantum-hard-rod model}

In the classical limit, for $\Gamma=0$, the ground state is ferromagnetic for $h<h_c=Jq/(q-1)$ and given in the form: $000000\dots$, $111111\dots$, $\dots$, $(q-1)(q-1)(q-1)(q-1)(q-1)(q-1)\dots$ having energy: $E_0^{(FM)}=-LJ-Lh/q$. For large alternating fields, $h>h_c$ we have an alternating ground state: $012\dots(q-1)012\dots(q-1)\dots$ with an energy $E_0^{(AL)}=-Lh$. At the critical end-point there is a first-order transition and the ground state is infinitely degenerate. The degenerate ground states can be obtained from the alternating ground state by replacing a string of consecutive sites of length $(q-1)$ and having states at the two neighbouring sites say $\tilde{s}$ by that state:
\be
\dots \tilde{s}\overbrace{s_1 s_2 \dots s_{q-1}}^{{\rm string}} \tilde{s} \dots \to \dots \tilde{s}\overbrace{\tilde{s}  \tilde{s}  \dots \tilde{s} }^{{\rm string}}  \tilde{s} \dots\;, 
\label{string}
\ee
in which $s_i=(\tilde{s} +i, {\rm mod}~q)$. Indeed, the energy of the string and the two neighbouring sites is $(q+1)h$ in the original state and it is $qJ+2h$ in the transformed state, which are equal at $h=h_c$. Consequently the typical form of a degenerate ground states is in the form:
\be
\begin{aligned}
\underbrace{sss\dots s}_{\rm altern}&\textcolor{red}{\overbrace{sss\dots s}^{\rm string}}\underbrace{sss\dots s}_{\rm altern}\textcolor{red}{\overbrace{sss\dots s}^{\rm string}}\dots\underbrace{sss\dots s}_{\rm altern}\\
\downarrow\downarrow\downarrow \dots \downarrow &\textcolor{red}{\uparrow} \downarrow\downarrow ... \downarrow \downarrow \downarrow\downarrow ... \downarrow \textcolor{red}{\uparrow}  \downarrow\downarrow ... \downarrow \dots \downarrow \downarrow\downarrow ... \downarrow
\label{stringstate}
\end{aligned}
\ee
in which the parts indicated by $altern$, are the same as in the alternating ground states, while the parts indicated by $string$ are constructed as explained in Eq.(\ref{string}). In the lower line this state is represented by a two-state basis, so that the leftmost site of a string is indicated by an up spin, while to all other sites down spins are assigned. Since the length of the strings is $(q-1)$ and the alternating parts contain at least one site two consecutive strings are separated at least by $q$ sites. In this way the degenerate ground states can be represented by a set of hard rods of length $q$, in which a rod consists of the sites of a string plus its left neighbouring site. 

The same type of hard rods of length $m$ are identified earlier in the degenerate ground state of the multispin-coupling model in Eq.(\ref{Hamilton_M}) at its classical end-point at $h_M=-J_Mm$.  As described in Ref.\cite{PhysRevB.40.2362} in the degenerate ground state the majority of spins are in $\downarrow$ states and the distance between two neighbouring $\uparrow$ spins is at least $m$. (This type of state is shown in the lower row of Eq.(\ref{stringstate}), with the correspondence $m=q$). In the vicinity of the CEP ($\Delta h_R=h_M+mJ_M \to 0$, $\Gamma_R \to 0$, but $\Gamma_R/\Delta h_R >0$) the effective Hamiltonian is given by\cite{PhysRevB.40.2362}:
\be
\begin{aligned}
&\hat{H}_R=-\Delta h_R \sum_{i} \sigma_i^z-\Gamma_R \sum_i \sigma_i^x\\  
&+J_1\sum_i(\sigma_i^z+1)(\sigma_{i+1}^z+1)  + J_2\sum_i(\sigma_i^z+1)(\sigma_{i+2}^z+1) + \dots
\end{aligned}
\label{rod}
\ee
where  $J_i=\infty$ for $i=1,2,\dots,m-1$ and it is zero for $i \ge m$.

Now turning back to the $q$-stae Potts model in an alternating field the degenerate ground state at the CEP is equivalent to that of the hard rod model with $q=m$ and at $\Gamma_R=0$. This correspondence is visualised in the lower row of Eq.(\ref{stringstate}). Leaving the CEP for $\Delta h=h_c-h \to 0^+$ and $\Gamma \to 0^+$ an effective model is obtained in degenerate perturbation expansion. The secular matrix has the same structure, as for the equivalent hard rod model: in leading order non-vanishing matrix-elements are between the corresponding states, see in Eq.(\ref{stringstate}). The only difference, however, that for the Potts chain a string of $(q-1)$ sites have to be flipped, which can be performed in the $(q-1)$th order of the perturbation calculation. This is the reason why the equation of the phase-boundary is given by: $\Delta h \sim \Gamma^{q-1}$ in the vicinity of the CEP, as discussed in Sec.\ref{sec:phase_diag}.

\bibliography{Pottsfield}

\end{document}